\documentstyle[12pt]{article}
\newtheorem{th}{Theorem}

\def\sar#1{\stackrel{#1}{\longrightarrow}}
\def\sd{\stackrel{def}{=}}
\def\sq{\stackrel{?}{=}}
\def\0{(0,0,0)}
\def\p#1#2#3#4{P^{(#1,#2)}_{x_{#1}^{#3}x_{#2}^{#4}}}
\def\t{transformation}

\def\proof{{\sl Proof }}
\def\be{\begin{equation}}
\def\ee#1{\label{#1}\end{equation}}
\def\a{\alpha}
\def\b{\beta}
\def\RD{Ribaucour-Darboux}
\def\R{Ribaucour}

\begin{document}

\author{Ganzha E.I.\\
  Dept. Mathematics, \\
  Krasnoyarsk State Pedagogical University \\
  Lebedevoi, 89, 660049, Krasnoyarsk, RUSSIA, \\
 e-mail: ganzha@edk.krasnoyarsk.su }
\title{On completeness of the Ribaucour \t s for triply orthogonal curvilinear
coordinate systems in ${\bf R}^3$\thanks{The research described in
this contribution was supported in part by
grant RFBR-DFG No 96-01-00050}}
\date{9 June 1996}
\maketitle

Let us consider the following system introduced in
 \cite{dar-ort} and describing triply orthogonal curvilinear
coordinate systems in ${\bf R}^3$:
\be
\left\{
\begin{array}{l}
 \frac{\partial \b}{\partial x_0} = \lambda^2
          \frac{\partial \a}{\partial x_0} \\
 \frac{\partial \b}{\partial x_1} = -\lambda^2
          \frac{\partial \a}{\partial x_1} \\
 \frac{\partial(\a + \b)}{\partial x_2} = (\b - \a)
          \frac{\partial \log \lambda}{\partial x_2}
\end{array}
\right.
\ee{ud}
If we will suppose that $\lambda(x_0,x_1,x_2)$ is a given function then
 (\ref{ud}) becomes a {\it linear} system for the unknowns
$\a(x_0,x_1,x_2)$, $\b(x_0,x_1,x_2)$. Its compatibility conditions
reduce to a single equation for the function
 $M= \log \lambda$ (which we will call hereafter a "potential"):
\be
\frac{\partial^3 M}{\partial x_0\partial x_1\partial x_2}
  = \mbox{cth}(M)\frac{\partial M}{\partial x_0}
        \frac{\partial M}{\partial x_1\partial x_2} +
 \mbox{th}(M)\frac{\partial M}{\partial x_1}
        \frac{\partial M}{\partial x_0\partial x_2}.
\ee{up}

The system (\ref{ud}) has a remarkable property
(discovered by G. Darboux, see \cite{dar-ort}):
if one has two its solutions $\{\a,
\b\}$ and $\{\overline\a, \overline\b\}$ for a given potential
$M(x_0,x_1,x_2)$ then one can obtain (performing a quadrature) a solution
 $\{\overline\a_1, \overline\b_1\}$ of (\ref{ud}) with a new potential.
The formulas of this transformation are given by:
\be
M_1 = - M_0 + \ln (\b /\a),
\ee{ip}
\be
\overline\a_1= \overline\a - \frac{\xi}{\b},
\quad \overline\b_1= \overline\b - \frac{\xi}{\a},
\ee{ir}
where $\xi(\overline\a,\overline\b, x_2)$ is obtained with a quadrature
from the following compatible system
\be
\frac{\partial \xi}{\partial \overline\a} = \b, \quad
\frac{\partial \xi}{\partial \overline\b} = \a, \quad
\frac{\partial \xi}{\partial x_2} = -\frac{(\overline\a+\overline\b)^2}{2}
     \frac{\partial}{\partial x_2}
    \left(\frac{\a -\b}{\overline\a -\overline\b}\right),
\ee{kv}
in which one is to suppose that
$\a$, $\b$ are expressed as functions of $\overline\a$,
$\overline\b$, $x_2$. Note that $\a_1 = 1/\overline\b$,
$\b_1 = 1/\overline\a$  also give a solution of $(M_1)$
 (as we will denote below the system (\ref{ud})  with the potential $M_1$),
and $\xi_1 = -\frac{\xi}{\a\b}$ is the corresponding to $\a_1$, $\b_1$,
$\overline\a_1$, $\overline\b_1$ solution of (\ref{kv}).
This \t\ is an analogue of the well known Moutard \t{}
 (see. \cite{darboux,bianchi,ganzha2})  and was given in the form (\ref{ud})
by G. Darboux in his monograph \cite{dar-ort}.
As Darboux had shown (\cite{dar-ort}),  (\ref{ip}),
(\ref{ir}), correspond to the \t\ of the connected to the pair
 $\{\a,\b\}$, $\{\overline\a,\overline\b\}$
of solutions of (\ref{ud}) triply orthogonal curvilinear
coordinate system by inversion. The variation of the second solution
 $\{\overline\a,\overline\b\}$ correspond to the so called Combescure
transformations of the orthogonal coordinate system;
composition of inversions and Combescure \t s are equivalent to
compositions of the \R\ \t s.
Hence we will call the \t\ (\ref{ip}),
(\ref{ir}) the \RD\ \t.

Let us construct starting from some fixed potential
 $M_0(x_0,x_1,x_2)$ for which one can find
 {\large all} solutions of (\ref{ud}) (e.g.
 $M_0={\mbox const}$) a chain of \RD\ \t s
 $(M_0) \rightarrow
(M_1) \rightarrow (M_2) \rightarrow \ldots\ (M_k) \rightarrow \ldots$.
The potential $M_k$ will depend (apriori) on
 $3k$ functions of 1 variable --- initial data for the solutions
$(\a_s, \b_s)$
of $(M_s)$, $s= 0, \ldots\ , k-1$. Indeed if we will define
$u=\log(\b/\a)$ then
$M_{s+1} = -M_{s} +u$ and the equations for $u$ are given
 by the elimination of
$\b$ from (\ref{ud}) with $\b=\a \exp(u)$:
\be
\begin{array}{l}
\partial_0\partial_1 u = -\frac{e^{4M} + e^{2u}}{e^{4M} - e^{2u}}
     \partial_1u\partial_0u +
    \frac{e^{2M} - e^{u}}{e^{2M} + e^{u}}
     \partial_1u\partial_0M +
    \frac{e^{2M} + e^{u}}{e^{2M} - e^{u}}
     \partial_0u\partial_1M, \\
\partial_0\partial_2 u = \frac{(e^{2M} - e^{u})(e^{u} - 1)}
       {e^u(e^{2M} +1)}
     \partial_0\partial_2M +
    \frac{2(e^{2M} + e^{2u})}{(e^{2M} - e^{u})(e^{u} +1)}
     \partial_0u\partial_2M -
    \frac{(e^{2M} + e^{2u})}{(e^{2M} - e^{u})(e^{u} +1)}
     \partial_0u\partial_2u, \\
\partial_1\partial_2 u = \frac{(e^{2M} + e^{u})(e^{u} - 1)}
       {e^u(e^{2M} -1)}
     \partial_1\partial_2M +
    \frac{2(e^{2M} - e^{2u})}{(e^{2M} + e^{u})(e^{u} +1)}
     \partial_1u\partial_2M -
    \frac{(e^{2M} - e^{2u})}{(e^{2M} + e^{u})(e^{u} +1)}
     \partial_1u\partial_2u,
\end{array}
\ee{uu}
$\partial_i = \partial /\partial x_i$.
As one can easily check,  (\ref{up}) give the necessary and sufficient
compatibility conditions for
(\ref{uu}). Then the known theorem by Darboux
 \cite[p. 335]{dar-ort} states that if one will fix in a neighborhood
of $(x_0^{(0)},x^{(0)}_1,x^{(0)}_2)$ the initial values $\varphi_0(x_0)=
u(x_0,x^{(0)}_1,x^{(0)}_2)$,  $\varphi_1(x_1)=
u(x_0^{(0)},x_1,x^{(0)}_2)$,  $\varphi_2(x_2)=
u(x_0^{(0)},x^{(0)}_1,x_2)$ (supposed to be continuously differentiable)
then there exists the unique solution of
(\ref{uu}) with such initial data.
So for the \RD\ \t s the potential
$M_{s+1}$ depends on $M_s$ and $3$ functions of $1$ variable
 --- the initial data for $u_s$.

As we have shown in \cite{ganzha1}, the necessary quadratures (\ref{kv})
may be avoided for
 $(M_s)$, $s \geq 2$ ("the B\"acklund cube"). For the aforementioned case of
the Moutard \t s the so called "pfaffian formulas" (\cite{ath-nim}) are known
as an analogue of the "wronskian formulas" for
 $(1+1)$-dimensional integrable systems (\cite{mat-sal}).

The theory of \R\ \t s was extensively studied in the monographs by Darboux
 \cite{dar-ort},  \cite{darboux} and in numerous papers of his contemporaries
 \cite{bianchi,tzitzeica,demoulin,guichard}
 for the general case of the $n$-dimensional euclidean space.
For example L.Bianchi constructed the traditional (~$(1+1)$-dimensional)
B\"acklund \t\ for the system of the resonant 3-wave interaction
using Egorov reduction of the equations of triply orthogonal curvilinear
coordinate systems (see \cite{bianchi,tsarev}).
One may ask how wide is the class of potentials for (\ref{ud})  obtainable (via
algebraic operations!) from a given potential $M_0$. In the theory of
$(1+1)$-dimensional integrable equations the question of density of the finite
gap solutions of the Korteweg-de Vries equation in the class of all
quasiperiodic functions was answered positively.
In  \cite{ganzha2} we have shown that for the case of the Moutard \t s
(such \t s possess the typical properties of $(2+1)$-dimensional B\"acklund \t
s while the Moutard equation itself $u_{xy} = M(x,y) \, u$ is formally
$(1+1)$-dimensional) the set of the potentials $M_k$ obtainable form any fixed
$M_0$ is "locally dense" in the space of all smooth functions in a sense to be
detailed below.

In this paper we solve the problem of (local) density of
 $M_k(x_0,x_1,x_2)$ obtainable from a given
initial $M_0(x_0,x_1,x_2)$ in the space of all solutions of (\ref{up}).
Hence we have proved that the \R\ \t\ lets us to construct "almost all"
orthogonal curvilinear coordinate system in
${\bf R}^3$, consequently the systems of equations describing such coordinate
systems ((\ref{up}) in particular) shall be considered as "true" integrable
$(2+1)$-dimensional  nonlinear systems.

Note that due to the known theorem by Darboux
\cite[p. 335]{dar-ort} the initial data for
 (\ref{up}) are given by $\Phi^{(0,1)}(x_0,x_1) =M(x_0,x_1,0)$,
$\Phi^{(0,2)}(x_0,x_2) =M(x_0,0,x_2)$, $\Phi^{(1,2)}(x_1,x_2) =M(0,x_1,x_2)$
in a neighborhood of $(x^{(0)}_0,x^{(0)}_1,x^{(0)}_2)=\0$.

\begin{th}
Let us fix some $M_0(x_0,x_1,x_2) \in C^\infty$
in a neighborhood of  $\0$. Then for any $N=0,1,2,\ldots $
one can find $K$ such that for an arbitrary triple system of numbers
$P^{(0,1)}_{z_1 \ldots\ z_k}$, $P^{(0,2)}_{t_1 \ldots\ t_k}$,
$P^{(1,2)}_{r_1 \ldots\ r_k}$, $z_i \in \{x_0,x_1\}$,
$t_i \in \{x_0,x_2\}$, $r_i \in \{x_1,x_2\}$, $0 \leq k \leq N$, the
corresponding derivatives of the potential $M_k$ from the chain of the \RD\ \t
s coincide with $P^{(i,j)}_{\ldots}$:
\be
\partial_{0}^p\partial_1^q M_K = \p01pq\sd
P^{(0,1)}_{\underbrace{x_0 \ldots\ x_0}_{p}\underbrace{x_1 \ldots\ x_1}_{q}},
\ee{poi}
\be
\partial_{0}^p\partial_2^q M_K = \p02pq\sd
P^{(0,2)}_{\underbrace{x_0 \ldots\ x_0}_{p}\underbrace{x_2 \ldots\ x_2}_{q}},
\ee{poii}
\be
\partial_{1}^p\partial_2^q M_K = \p12pq\sd
P^{(1,2)}_{\underbrace{x_1 \ldots\ x_1}_{p}\underbrace{x_2 \ldots\ x_2}_{q}},
\ee{piii}
\end{th}
Obviously we suppose
$$
P \sd P^{(0,1)}=P^{(0,2)}=P^{(1,2)}=M\0,\quad P^{(0)}_{x_0^i}\sd
P^{(0,1)}_{x_0^i}=P^{(0,2)}_{x_0^i}=\partial_0^iM_K\0,
$$
$$
 P^{(1)}_{x_1^i}\sd P^{(0,1)}_{x_1^i}=P^{(1,2)}_{x_1^i}=\partial_1^iM_K\0,
\quad
 P^{(2)}_{x_2^i}\sd P^{(0,2)}_{x_2^i}=P^{(1,2)}_{x_2^i}=\partial_2^iM_K\0.
$$

\proof will be given inductively. For $N=0$ choose $K=1$; from (\ref{ip})
$M_1\0= -M_0\0 +u_0\0$. Since (for a given $M_0$) one may set
the initial value $u_0\0$ to be arbitrary, and all
$\partial_i^ku_0\0$ set to be $0$ (the initial Goursat data for (\ref{uu}),
one can easily show (differentiating $M_1= -M_0+u_0$)
that for $N=0$, $K=1$, the following  {\bf basic induction proposition} holds:

\noindent {\sl  for any $N$ one can find such $K$ that
the corresponding derivatives
$\partial_i^m\partial_j^n M_K$, $i,j \in \{0,1,2\}$, $m+n \leq N$, of a
conveniently chosen $M_K$ in
$\0$ coincide with the given
$\p ijmn$ for  $m+n \leq N$: $\p ijmn=\partial_i^m\partial_j^n M_K\0$.
Additionally the higher derivatives
$\partial_i^m\partial_j^n M_K$, $m+n > N$,
depend only on $\p klpq$, $p+q \leq N$, in the following way:
$\partial_i^m M_K$, $m>N$, coincide with $\partial_i^m M_0$ 
(they do not depend
on the lower derivatives $\partial_i^m\partial_j^n M_K$, $m+n \leq N$);
$\partial_0^m\partial_1^n M_K$, $m+n > N$, $m,n >0$,
 depend LINEARLY only upon
$\p01pq=\partial_0^p\partial_1^q M_K$, $p+q \leq N$, $p \leq m$;
$\partial_0^m\partial_2^n M_K$,
$\partial_1^m\partial_2^n M_K$, $m+n > N$, $m,n >0$, depend upon
$\p01pq=\partial_0^p\partial_1^q M_K$, $p+q \leq N$, as well as upon
 $\p02ks$, $\p12ks$, for $k+s \leq N$, $s \leq n$.
}

{\bf Inductive step}.
Suppose that the basic induction proposition is true for
the derivatives of $M_K$ of orders $\leq N=N_0$, let us prove it for $N=N_0+1$.
Let  $K_0$ be the corresponding to
$N=N_0$ number of $M_{K_0}$.
 Performing $3N_0+3$ consecutive \RD\ \t s
$M_{K_0} \sar{u=u_1}M_{K_0+1} \sar{u=u_2} \ldots  \sar{u=u_{N_0}}
M_{K_0+N_0} \sar{u=v_1} M_{K_0+N_0+1} \sar{u=v_2}  \ldots  \sar{u=v_{N_0}}
M_{K_0+2N_0} \sar{u=w_1} M_{K_0+2N_0+1} \sar{u=w_2}  \ldots  \sar{u=w_{N_0}}
M_{K_0+3N_0} \sar{u=u^{(0)}} M_{K_0+3N_0+1} \sar{u=u^{(1)}} M_{K_0+3N_0+2}
  \sar{u=u^{(2)}}  M_{K_0+3N_0+3}$ and denoting for simplicity
$Q(x_0,x_1,x_2)=M_{K_0+3N_0+3}$ we have
\be
Q= (-1)^{N_0+1}\Big(M_{K_0} -u_1+u_2- \ldots \pm w_{N_0}\mp u^{(0)}
\pm u^{(1)} \mp u^{(2)} \Big).
\ee{uku}
Let us call the {\it principal} derivative of the function
 $u_i$, $1 \leq i \leq N_0$, its derivative
$\partial_0^i u_i$ at $\0$, and its   {\it auxiliary} derivative the
derivative
$\partial_1^{N_0+1-i} u_i$ at the same point.
 For $v_i$ and $w_i$, $ 1\leq i \leq
N_0$, respectively we suppose
$\partial_2^{i} v_i$, $\partial_2^{i} w_i$
to be the principal and
$\partial_0^{N_0+1-i} v_i$, $\partial_1^{N_0+1-i} w_i$,
to be the auxiliary derivatives at $\0$.
All the other (nonmixed) derivatives of arbitrary orders for these functions
will be set to $0$ at $\0$. For $u^{(0)}$,
$u^{(1)}$, $u^{(2)}$ we suppose $\partial_0^{N_0+1}u^{(0)}$,
$\partial_1^{N_0+1}u^{(1)}$, $\partial_2^{N_0+1}u^{(2)}$ at $\0$ to be the
principal derivatives,
all the other (nonmixed) derivatives of arbitrary orders for these functions
are set to $0$ at $\0$ (no auxiliary derivatives).
Let us fix the values of $u_i$, $v_i$, $w_i$, $u^{(k)}$
and the values of their auxiliary derivatives at $\0$
to be "generic"; more precisely the respective inequalities for them
will be specified below, such inequalities may be obtained using the values of
$\p ijks$, $k+s \leq N_0+1$.

Inside the given step of the main induction
 (for $N=N_0+1$) we will perform an auxiliary induction over
 $m$ in order to show the validity of the main inductive proposition
for $\partial_0^m\partial_1^n Q$ for all $m$, $n$ for conveniently chosen
principal derivatives of $u_i$, $u^{(0)}$, $u^{(1)}$.

First of all for $m=0$ $\partial_1^{N_0+1}Q =
(-1)^{N_0+1}\Big( \partial_1^{N_0+1}M_{K_0} \pm
 \partial_1^{N_0+1}u^{(1)}\Big)$. The value of
$ \partial_1^{N_0+1}M_{K_0} $ at $\0$ as we know from the main induction
proposition is equal to
 $ \partial_1^{N_0+1}M_{0}$. Consequently one may unambiguously determine
the principal derivative $ \partial_1^{N_0+1}u^{(1)}$ in such a way that
$ \partial_1^{N_0+1}Q = P^{(1)}_{x_1^{N_0+1}}$. Since
 $\partial_1^{n}Q =
(-1)^{N_0+1}\Big( \partial_1^{n}M_{K_0} \pm\partial_1^n u_{N_0+1-n} \pm
 \partial_1^{n}w_{N_0+1-n}\Big)$, $0 < n \leq N_0$, one may uniquely determine
the derivatives  $\partial_1^n M_{K_0}$ (they may be chosen arbitrarily
according to the same proposition) using the fixed above
auxiliary derivatives of $u_i$, $v_i$,  $w_i$ in such a way that
 $\partial_1^n Q$ be equal to the desired value
 $P^{(1)}_{x_1^n}$. Note that due to
\be
M_{K_0+s} = (-1)^s \Big(M_{K_0} -u_1 + \ldots \Big)
\ee{ums}
the values of $M_{K_0+s}$ and all their nonmixed derivatives at the origin
may be consecutively (using induction on $s$) found from
the derivatives of $M_{K_0}$ and $u_i$,
$v_i$, $w_i$, $u^{(s)}$ with respect to $x_=1$.
 So they depend only on $\p01mn$. Applying $\partial_1^n$, $n > N_0+1$,
to (\ref{uku}) we see that $\partial_1^n Q \equiv \partial_1^n M_{K_0} \equiv 
=\partial_1^n M_0$ due to the choice of the nonmixed derivatives of 
$v_i$, $w_i$, $u^{(s)}$ and the inductive proposition.

The equality $P^{(0)}_{x_0^{N_0+1}}= \partial_0^{N_0+1}Q$ is easy to obtain as
before (for $P^{(1)}_{x_1^{N_0+1}}= \partial_1^{N_0+1}Q$)
choosing the appropriate principal derivative
 $\partial_0^{N_0+1}u^{(0)}$. The choice of the lower derivatives 
$\partial_1^n M_0$, $n \leq N_0$ is analogous to the above case.
The proposition about the independence of the higher derivatives on the lower
ones is obvious as well.

Let us choose the principal derivative $\partial_2^{N_0+1} u^{(0)} \0$
in such a way that $P^{(2)}_{x_2^{N_0+1}} = \partial_2^{N_0+1}Q$. The
required dependence of $\partial_2^mQ$, $m > N_0+1$ is obvious.

Thus for  $m=0$ and $m=N_0+1$ we have:
\begin{itemize}
\item[\rm a)] the principal derivatives
 $\partial_0^{N_0+1}u^{(0)}$,  $\partial_1^{N_0+1}u^{(1)}$ and
 $\partial_1^iu_i$, $i \leq m$, are already chosen in such a way that
the respective equalities (\ref{poi})
$p+q \leq N_0+1$, $p \leq m$, are achieved (for $Q=M_K$);
\item[\rm b)] the higher derivatives
$ \partial_0^p\partial^{q}_1 Q$, $p+q > N_0+1$, $p \leq m$,
and the already defined principal derivatives
depend (according to the construction of $Q$) only on the values of
$\p01st$, $s+t \leq N_0+1$, $s \leq m$.
\end{itemize}

We will show the validity of a), b) for  $0<m<N_0+1$
 by an auxiliary induction
over $m$, $m \leq N_0$. In order to perform its step
($m=m_0+1$) apply $\partial_0^{m_0+1}\partial_1^{N_0-m_0}$
to (\ref{uku}) taking into consideration the equations (\ref{uu}),
(\ref{ums}). We will obtain an expression including in its right hand side
(besides the already defined values of the principal derivatives and
$\partial_0^p\partial_1^qM_{K_0}$, $p \leq m_0$)
the values of $\partial_0^{m_0+1}\partial_1^{N_0-m_0}M_{K_0}$,
$\partial_0^{m_0+1}\partial_1^qM_{K_0}$, $q < N_0-m_0$,
as well as the only (so far undefined) principal derivative
 $\partial_0^{m_0+1}u_{m_0+1}\0$ with the coefficient
$$
c_{m_0+1} = - \frac{\exp(4M_L) + \exp(2 u_{m_0+1})}
{\exp(4M_L) - \exp(2 u_{m_0+1})}\partial_1^{N_0-m_0}u_{m_0+1} + F,
$$
$L=K_0+m_0+1$, where $F$ comprises only the defined above values of
 $u_i$, $v_i$,$w_i$, their auxiliary derivatives at $\0$, the
defined during the previous steps of the auxiliary induction
principal derivatives $\partial_0u^i$, $i \leq m_0$, $i \neq m_0-1$,
as well as
$\partial_1^q M_{K_0}\0$ with $q\leq N_0$.

We need to show that
$\partial_0^{m_0+1}\partial_1^qM_{K_0}$, $q < N_0-m_0$,
and the principal derivative  $\partial_0^{m_0+1}u_{m_0+1}\0$
may be determined uniquely by the equality to be achieved and by
(\ref{poi}) with $p+q \leq N_0+1$, $p \leq m_0+1$: due to $q < N_0-m_0$
the coefficient of the principal derivative in
$\partial_0^{m_0+1}\partial_1^qQ$   includes only the initial values
of the functions at $\0$ and $\partial_1^qM_{K_0}$ which were determined above.
We remark now that the value of
  $\partial_0^{m_0+1}\partial_1^rM_{K_0+s}\0$ depend linearly on the
values of $\partial_0^{m_0+1}\partial_1^qM_{K}$ and
 $\partial_0^{m_0+1}u_{m_0+1}\0$.
 Expressing
the values of $\partial_0^{m_0+1}\partial_1^qM_{K_0}$
linearly via the aforementioned quantities
and substituting into the equality we are to establish on this inductive step,
we may use the condition of genericity of
$u_{m_0+1}\0$ and its auxiliary derivative
 $\partial_1^{N_0-m_0}u_{m_0+1}$ and conclude that the resulting coefficient
of the principal derivative  is not equal to
 $0$ (this is one of the inequalities constraining them) and consequently
we can determine the principal derivative
 $\partial_0^{m_0+1}u_{m_0+1}\0$
and afterwards
$\partial_0^{m_0+1}\partial_1^qM_{K_0}$, $p\leq N_0-m_0$,
uniquely in such a way that
$\p01{m_0+1}{N_0-m_0} = \partial_0^{m_0+1}\partial_1^{N_0-m_0} Q$.
The proposition about the dependence of
$ \partial_0^{m_0+1}\partial_1^{q} Q$, $m_0+1+q >N_0+1$, only upon
the used $\p01st$, $s \leq m_0+1$, can be easily obtained from
(\ref{uku}) taking into consideration (\ref{uu}) and (\ref{ums}). The auxiliary
induction is therefore completed.

Consider now the equations $$\p02{N_0+1-m}{m} =
\partial_0^{N_0+1-m}\partial_2^mQ\0,$$ $$\p12{N_0+1-m}{m}   =
\partial_1^{N_0+1-m}\partial_2^mQ\0$$ simultaneously.
We will show the possibility to satisfy them by an appropriate choice of the
principal derivatives
$\partial_2^i v_i$, $\partial_2^i w_i$ performing the second auxiliary
induction over $m$.

Indeed for $m=0$ the above equations are already true:
$P^{(0)}_{x_0^{N_0+1}}=\p02{N_0+1}0=\p01{N_0+1}0$,
$P^{(1)}_{x_0^{N_0+1}}=\p12{N_0+1}0=\p010{N_0+1}$, as we established above.
The proposition about the dependence of the
 higher order derivatives is trivial in this case.
Suppose all these facts to be true for $m \leq m_0$, we will show their
validity for $m= m_0+1$. We have (cf. (\ref{uu}), (\ref{ums}))
\be
\left\{
\begin{array}{l}
\p02{N_0-m_0}{m_0+1} \sq \partial_0^{N_0-m_0}\partial_2^{m_0+1}Q =
  (-1)^{N_0+1} \Big(k_1 \partial_0^{N_0-m_0}\partial_2^{m_0+1}M_{K_0} \\
\qquad
+a \partial_2^{m_0+1}v_{m_0+1}+b \partial_2^{m_0+1}w_{m_0+1} \Big) + F, \\
\p12{N_0-m_0}{m_0+1} \sq \partial_1^{N_0-m_0}\partial_2^{m_0+1}Q =
  (-1)^{N_0+1} \Big(k_2 \partial_1^{N_0-m_0}\partial_2^{m_0+1}M_{K_0} \\
\qquad
+c \partial_2^{m_0+1}v_{m_0+1}+d \partial_2^{m_0+1}w_{m_0+1} \Big) + G,
\end{array}
\right.
\ee{ufz}
where  $F$ and $G$ comprise all the terms including only the already found
during the previous inductive steps quantities as well as
 $\partial_0^{s}\partial_2^{m_0+1}M_{K_0}$, $s < N_0-m_0$,
 $\partial_1^{s}\partial_2^{m_0+1}M_{K_0}$, $s < N_0-m_0$.
The coefficients $k_i$ include
only the values  $u_i$, $v_i$, $w_i$, $u^{(i)}$ at $\0$. The coefficients
 $a$, $d$ include respectively the auxiliary derivatives
$\partial_0^{N_0-m_0}v_{m_0+1}$, $\partial_1^{N_0-m_0}w_{m_0+1}$
with the coefficients $-(\exp(2M_L) + \exp(2v_{m_0+1}))/\big(
(\exp(2M_L) - \exp(2v_{m_0+1}))( \exp(2v_{m_0+1})+1)\big)$,
$-(\exp(2M_T) - \exp(2w_{m_0+1}))/\big(
(\exp(2M_T) + \exp(2w_{m_0+1}))( \exp(2w_{m_0+1})+1)\big)$, $L=K_0+N_0+m_0+1$,
 $T=K_0+2N_0+m_0+1$, definable via the fixed quantities, these coefficient do
not vanish due to the genericity condition (one can find the corresponding
inequalities from $\p ijks$).
Again from the genericity of the auxiliary derivatives one can suppose
the determinant $\left|\begin{array}{cc}
a & b \\
c & d
\end{array}
\right|$
to be nonzero at $\0$ (again the inequalities for the corresponding auxiliary
derivatives expressible in terms of
$\p ijks$). $\partial_0^{N_0-m_0}\partial_2^{m_0+1}M_{K_0}$,
$\partial_1^{N_0-m_0}\partial_2^{m_0+1}M_{K_0}$ according to the main inductive
proposition may be found from
$\p01pq$, $\p02st$, $\p12st$, $s \leq m_0+1$.

The quantities $\partial_0^{s}\partial_2^{m_0+1}M_{K_0}$, $s < N_0-m_0$,
 $\partial_1^{s}\partial_2^{m_0+1}M_{K_0}$, $s < N_0-m_0$,
may be found from (\ref{poii}), (\ref{piii}) with $q=m_0+1$. Indeed
applying $\partial_0^{s}\partial_2^{m_0+1}$, $s < N_0-m_0$,
$\partial_1^{s}\partial_2^{m_0+1}$, $s < N_0-m_0$, to  (\ref{uku}),
one can conclude from (\ref{uu})
that the coefficients of the undefined so far principal derivatives
$\partial_2^{m_0+1}v_{m_0+1}$, $\partial_2^{m_0+1}w_{m_0+1}$ due to
$s < N_0-m_0$ vanish, consequently we (via an induction over $s$)
unambiguously find
 $\partial_0^{s}\partial_2^{m_0+1}M_{K_0}$, $s < N_0-m_0$,
 $\partial_1^{s}\partial_2^{m_0+1}M_{K_0}$, $s < N_0-m_0$
(their coefficients do not vanish due to the genericity condition of
the values of $u_i$ and the other functions at $\0$) in such a way that
 (\ref{poii}), (\ref{piii}) hold for
$q=m_0+1$, $p+q < N_0+1$.

Hence by an appropriate (and unique) choice of the principal derivatives
$\partial_2^{m_0+1}v_{m_0+1}$, $ \partial_2^{m_0+1}w_{m_0+1}$ one can satisfy
the equalities marked by the question sign in (\ref{ufz}).
Applying to (\ref{uku}) the operators $\partial_0^m\partial_2^n$,
$\partial_1^m\partial_2^n$, $m+n > N_0+1$, $n \leq m_0+1$, we get the necessary
result on the dependence of the higher derivatives
$\partial_0^m\partial_2^nQ$, $\partial_1^m\partial_2^nQ$ upon $\p ijks$.

Consequently the main inductive
proposition and the Theorem are proved.

{\sl Remark}. One may consider of interest to generalize this result for the
case of the \R\ \t s of the orthogonal curvilinear coordinate systems in the
$n$-dimensional euclidean space (\cite{dar-ort,bianchi}).
We shall note that the system describing such coordinates is an overdetermined
system with the initial (Goursat like) data given by
 $n(n-1)/2$ functions of {\large two} variables for arbitrary $n$.
Consequently we have an
 $(2+1)$-dimensional system possessing a
$(2+1)$-dimensional B\"acklund \t\ --- the \R\ \t. Unfortunately no
representation analogous to (\ref{ud})
is known for this case. Nevertheless one may conjecture the hypothesis about
the completeness of the \R\ \t s also in the $n$-dimensional case.

\end{document}